\documentclass[aps,prl,twocolumn]{revtex4}
\usepackage{graphicx}
\begin{document}


\title{Edge Excitations and Non-Abelian Statistics
in the Moore-Read State: A Numerical Study in the Presence of Coulomb
Interaction and Edge Confinement}

\author{Xin Wan$^1$}
\author{Kun Yang$^{2,1}$}
\author{E. H. Rezayi$^3$}

\affiliation{$^1$Zhejiang Institute of Modern Physics, Zhejiang
University, Hangzhou 310027, P.R. China} \affiliation{$^2$NHMFL and
Department of Physics, Florida State University, Tallahasse, Florida
32306, USA} \affiliation{$^3$Physics Department California State
University Los Angeles, Los Angeles, California 90032, USA}

\date{\today}

\begin{abstract}
We study the ground state and low-energy excitations of fractional
quantum Hall systems on a disk at filling fraction $\nu = 5/2$, with
Coulomb interaction and background confining potential. We find the
Moore-Read ground state is stable within a finite but narrow window
in parameter space. The corresponding low-energy excitations contain
a fermionic branch and a bosonic branch, with widely different
velocities. A short-range repulsive potential can stabilize a charge
$+e/4$ quasihole at the center, leading to a different edge
excitation spectrum due to the change of boundary conditions for
Majorana fermions, clearly indicating the non-Abelian nature of the
quasihole.
\end{abstract}

\maketitle

Fractional quantum Hall (FQH) liquids represent novel states of
matter with non-trivial topological order~\cite{wen95}, whose
consequences include chiral edge excitations and fractionally
charged bulk quasiparticles that obey Abelian or non-Abelian
fractional statistics. It has been proposed that the non-Abelian
quasiparticles can be used for quantum information storage and
processing in an intrinsically fault-tolerant
fashion~\cite{kitaev03,freedman02}, in which information is stored
by the degenerate ground states in the presence of these non-Abelian
quasiparticles, and unitary transformations in this Hilbert space
can be performed by braiding the
quasiparticles~\cite{dassarma05,bonesteel05}. While many Abelian FQH
states have been observed and studied in detail\cite{wen95}, thus
far there have been relatively few candidates for the non-Abelian
ones. The most promising candidate is the FQH state at Landau level
filling fraction $\nu = 5/2$\cite{willett87}. The leading candidate
for the ground state of this system is the Moore-Read (MR) paired
state~\cite{moore91}, which has been shown~\cite{moore91,nayak96} to
support fractionally charged, non-Abelian quasiparticles. The MR
state received strong support from numerical studies using sphere or
torus geometries~\cite{morf}. It has been proposed that the
non-Abelian nature of the quasiparticles in the MR state may be
detected through interference experiments in edge
transport~\cite{fradkin98,stern06}. In order to have a quantitative
understanding of the edge physics however, one needs to study the
interplay between electron-electron interaction and confining
potential, which may lead to edge structures that are more
complicated than those predicted by the simplest
theory~\cite{chklovskii}. This was found to be the case rather
generically in the FQH regime~\cite{wan03}.

In anticipation of experimental studies, in this work we perform
detailed numerical studies of edge excitations in the 5/2 FQH state
in finite-size systems with disc geometry, taking into account the
inter-electron Coulomb interaction and a semi-realistic model of the
confining potential due to neutralizing background charge. For a
limited parameter space, we find the ground state has substantial
overlap with the MR state. Within this parameter space we identify
the existence of chiral fermionic and bosonic edge modes, in
agreement with previous prediction. We find the fermionic mode
velocity is much lower than that of the bosonic mode. With suitable
short-range repulsive potential at the center, we show that a charge
$+e/4$ quasihole can be localized at the center of the system, and
its presence changes the spectrum of the fermionic edge mode. This
confirms the existence and non-Abelian nature of such fractionally
charged quasiparticles.

{\it The microscopic model.} We consider a microscopic model of a
two-dimensional electron gas (2DEG) confined to a two-dimensional
disk, with neutralizing background charge distributed uniformly on a
parallel disk of radius $a$ at a distance $d$ above the 2DEG. This
distance parameterizes the strength of the confining potential,
which decreases with increasing $d$. For $\nu = 5/2$, we explicitly
keep the electronic states in the first Landau level (1LL) only,
while neglecting the spin up and down electrons in the lowest Landau
level (0LL), assuming they are inert. The amount of positive
background charge is chosen to be equal to that of the half-filled
1LL, so the system is neutral. The choice of $a = \sqrt{4N}$, in
units of $l_{\rm B}$ (magnetic length), guarantees that the disk
encloses exactly $2N$ magnetic flux quanta for $N=2P$, corresponding
to $\nu = 1/2$ in the 1LL~\cite{note}. The rotationally invariant
confining potential comes from the  Coulomb attraction between the
background charge and the electrons. Using the symmetric gauge, we
can write down the following Hamiltonian for the electrons confined
to the 1LL:
\begin{equation}
\label{eqn:chamiltonian} H_{\rm C} = {1\over 2}\sum_{mnl}V_{mn}^l
c_{m+l}^\dagger c_n^\dagger c_{n+l}c_m +\sum_m U_mc_m^\dagger c_m,
\end{equation}
where $c_m^\dagger$ is the electron creation operator for the 1LL
single electron state with angular momentum $m$, $V_{mn}^l$'s are
the corresponding matrix elements of Coulomb interaction for the
symmetric gauge, and $U_m$'s are the matrix elements of the
confining potential.

\begin{figure}
\includegraphics[width=0.45\textwidth]{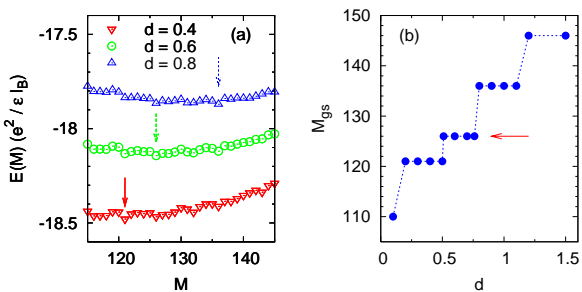}
\caption{ \label{fig:M_gs} (Color online) (a) Ground state energy
$E(M)$ in each angular momentum $M$ subspace for $N = 12$ electrons
with 22 orbitals in the 1LL (corresponding to $\nu = 1/2$). The
positive background charge is at a distance $d = 0.4$, 0.6, and 0.8,
in units of $l_{\rm B}$, above the electron plane. The total angular
momentum of the global ground state $M_{\rm gs}$ (indicated by
arrows) increases from 121, 126, to 136, respectively. The global
ground state at $d = 0.6$ has the same total angular momentum
$M_{\rm MR} = N(2N-3)/2 = 126$ as the corresponding $N = 12$ MR
state. The overlap between the two states is 0.47. The curves for $d
= 0.4$ and 0.8 have been shifted verically by 0.6 and -0.6,
respectively. (b) The total angular momentum of the global ground
state $M_{\rm gs}$ as a function of $d$. The arrow indicates the
plateau at which $M_{\rm gs} = M_{\rm MR}$. }
\end{figure}

{\it MR ground state.} We diagonalize the Hamiltonian
[Eq.~(\ref{eqn:chamiltonian})] for each Hilbert subspace with total
angular momentum $M$, and obtain the ground state energy $E(M)$.
Figure~\ref{fig:M_gs}(a) shows $E(M)$ vs $M$ for $N = 12$ electrons
in the 1LL in 22 orbitals, which is the minimum to accommodate the
corresponding MR state. As illustrated in Fig.~\ref{fig:M_gs}(b),
$M_{\rm gs}$ increases with increasing $d$. This is very similar to
what happens in the 0LL~\cite{wan03}, and reflects the interplay
between electron-electron Coulomb repulsion and confining potential;
as the confining potential weakens with increasing $d$, electrons
tend to move outward, resulting in bigger $M_{\rm gs}$. $M_{\rm gs}$
coincides with the total angular momentum of the $N = 12$ MR state
$M_{\rm MR} = N(2N-3)/2 = 126$ only within a small window: $0.51 \le
d/l_{\rm B} \le 0.76$. This contrasts with the situation for a
Laughlin filling fraction $\nu = 1/3$, where $M_{\rm
gs}=N(N-1)/2\nu$ (same as the corresponding Laughlin state) for a
substantially bigger window $d < d_c \approx 1.5 l_{\rm
B}$~\cite{wan03}. When the global ground state has the same total
angular momentum as the MR state, the overlap $|\langle \Psi_{\rm
gs}|\Psi_{\rm MR}\rangle|^2$ between the two is about 0.47. We note
that, in the absence of confining potential, the corresponding
overlaps are 0.46 and 0.45 for $N = 12$ and $14$, respectively.
These values are quite substantial, given that the size of Hilbert
subspaces are 16,660 and 194,668. But they are well below the
overlap ($> 0.95$) in the case of the Laughlin filling $\nu =
1/3$~\cite{wan03}. The reduced overlap and window for the MR state
reflect the fact that the paired state is much weaker compared to
the Laughlin state. The overlap is, however, quite sensitive to
small changes of system parameters; for example, we can increase the
overlap to above 0.7 for $N = 12$, by choosing $a = \sqrt{4N - 4}$,
and a change in the $V_1$ pseudopotential, $\delta V_1 = 0.03$. Such
sensitivity suggests that the MR state is rather ``fragile",
consistent with experiments at $\nu=5/2$.

{\it Edge excitations.} The MR state has non-trivial topological
order. While our numerical results indicate that the ground state
has substantial overlap with the MR state for properly chosen system
parameters, it does not directly reflect the topological order of
the system. One way to probe the topological order is to study edge
excitations, which is also of vital experimental importance. For
comparison, the Laughlin state supports one bosonic branch of chiral
edge excitations, whose properties have been studied in tunneling
experiments~\cite{chang03}. For $\nu = 5/2$, a neutral fermionic
branch of excitations has been predicted in addition to a bosonic
branch~\cite{wen95,milovanovic96}. The existence of both branches
makes the low-energy excitation spectrum of a microscopic model at
$\nu = 5/2$ richer, and their experimental consequences more
interesting~\cite{fendley}.

\begin{figure}
\includegraphics[width=0.45\textwidth]{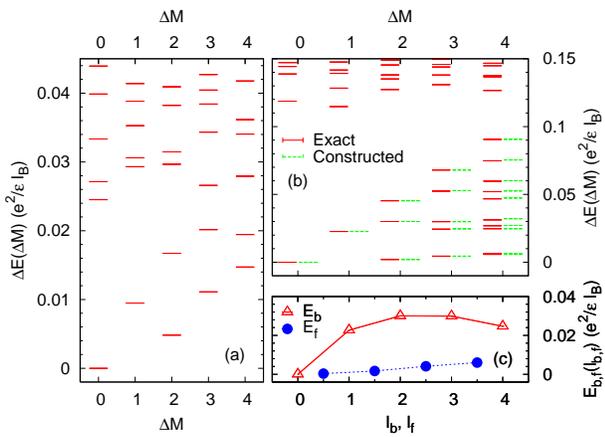}
\caption{ \label{fig:modes} (Color online) Low-energy excitations
$\Delta E(\Delta M)$ from exact diagonalization (solid lines) for $N
= 12$ electrons in 26 orbitals in the 1LL (corresponding to $\nu =
1/2$) for (a) Coulomb Hamiltonian [Eq.~(\ref{eqn:chamiltonian})] and
(b) mixed Hamiltonian [Eq.~(\ref{eqn:mixedhamiltonian}) with
$\lambda = 0.5$]. The neutralizing background charge for the Coulomb
part is deposited at $d = 0.6 l_{\rm B}$ above the electron plane.
(c) Dispersion curves of bosonic ($E_{\rm b}$) and fermionic modes
($E_{\rm f}$) of the system. These energies can be used to construct
the complete edge spectrum for the 12-electron system up to $\Delta
M = 4$ [dashed bars in (b)]. }
\end{figure}

Figure~\ref{fig:modes}(a) shows the low-energy excitations for pure
Coulomb interaction and the confining potential with $d = 0.6$ for
12 electrons in 26 orbitals. Apparently, there is no clear
distinction between fermionic and bosonic edge modes as well as bulk
modes, due to the relatively small bulk gap, and system size. The
situation here is similar to a related study on a rotating Bose
gas~\cite{cazalilla05}, and will be analyzed in detail elsewhere.
Here we focus instead on a model with mixed Coulomb interaction and
3-body interaction for clarity. The 3-body interaction alone
generates the MR state as its exact ground state with the smallest
total angular momentum. The mixed Hamiltonian is
\begin{eqnarray}
\label{eqn:mixedhamiltonian} H &=& (1 - \lambda) H_{\rm C} + \lambda
H_{3B}, \\
H_{3B} &=& - \sum_{i < j <
k}S_{ijk}[\nabla^2_i\nabla^4_j \delta({\bf r}_i - {\bf r}_j)
\delta({\bf r}_i -{\bf r}_k)],
\end{eqnarray}
where $S$ is a symmetrizer:
$S_{123}[f_{123}]=f_{123}+f_{231}+f_{312}$. We measure energies
in units of $e^2/\epsilon l_B$. As we will see, the
mixed interaction, which enhances the bulk excitation gap with
respect to that of edge excitations, allows for a clear separation
between the two in a finite system, effectively increasing the
system size.

Figure~\ref{fig:modes}(b) shows the low-energy excitations $\Delta
E(\Delta M)$ for 12 electrons in 26 orbitals in the 1LL for the
mixed Hamiltonian with $\lambda = 0.5$ and $d = 0.6 l_{\rm B}$.
There is a clear separation of the spectrum around $\Delta E = 0.1$,
below which we identify as edge modes. The total numbers of these
states are 1, 1, 3, 5, and 10 for $\Delta M = 0$-4, which agree with
the numbers of edge states expected for the MR
state~\cite{milovanovic96}. Notably, the lowest two levels for
$\Delta M = 4$ lie very close to each other.

In this case, we can further separate the fermionic and bosonic
branches of the edge states. The procedure is similar to but more
complicated than the one we used~\cite{wan03} to identify edge modes
in the Laughlin case at $\nu = 1/3$, where there is only one bosonic
branch of edge modes. The basic idea is to label the low-lying
states by two sets of occupation numbers $\{n_{\rm b}(l_{\rm b})\}$
and $\{n_{\rm f}(l_{\rm f})\}$ for bosonic and fermionic modes with
angular momentum $l_{\rm b,f}$, respectively. Since the fermionic
edge excitations are Majorana fermions that obey antiperiodic
boundary conditions~\cite{milovanovic96}, $l_{\rm f}$ must be
positive half integers and $\sum_{l_{\rm f}} n_{\rm f}(l_{\rm f})$
an even integer because fermion modes are occupied in pairs. The
angular momentum and energy of the state, measured from those of the
ground state, are $\Delta M = \sum_{l_{\rm b}} n_{\rm b}(l_{\rm b})
l_{\rm b} + \sum_{l_{\rm f}} n_{\rm f}(l_{\rm f}) l_{\rm f}$ and $
\Delta E = \sum_{l_{\rm b}} n_{\rm b}(l_{\rm b}) E_{\rm b}(l_{\rm
b}) + \sum_{l_{\rm f}} n_{\rm f}(l_{\rm f}) E_{\rm f}(l_{\rm f})$.
The difficulty here is, in addition to the bosonic modes, we also
have fermionic modes, and thus the convolution of fermionic and
bosonic modes. Fortunately, we note that the number of states with
nearly zero energy coincides with the number of fermionic edge
states expected by theory. Accordingly in our construction we will
assume these energies to have been evolved from combining two
Majorana fermions. Through careful analysis of the low-energy
excitations up to $\Delta M = 4$, we obtain the results of bosonic
and fermionic mode energies up to $l_{\rm b} = 4$ and $l_{\rm f} =
7/2$, respectively, plotted in Fig.~\ref{fig:modes}(c). The detailed
analysis will be published elsewhere. Using these 8 energies
[excluding the trivial $E_{\rm b}(0) = 0$], we can construct the
whole low-energy spectrum of the system up to $\Delta M = 4$, a
total of 20 states. The excellent agreement [see
Fig.~\ref{fig:modes}(b)] justifies our analysis, and thus our
result, that, energetically, fermionic modes are well separated from
bosonic modes. In contrast to the roughly linear dispersion of the
fermionic branch, the energy of the bosonic branch bends down
(despite a much bigger initial slope or higher velocity), suggesting
a potential vulnerability to edge reconstruction in the bosonic
branch~\cite{wan03}. These are not surprising since the bosonic
modes are charged; as a result its velocity is dominated by the
long-range nature of the Coulomb interaction in the long-wavelength limit,
but in the meantime it is also more sensitive to the competition
between Coulomb interaction and confining potential which can lead
to instability at shorter wavelength.

\begin{figure}
\includegraphics[width=0.45\textwidth]{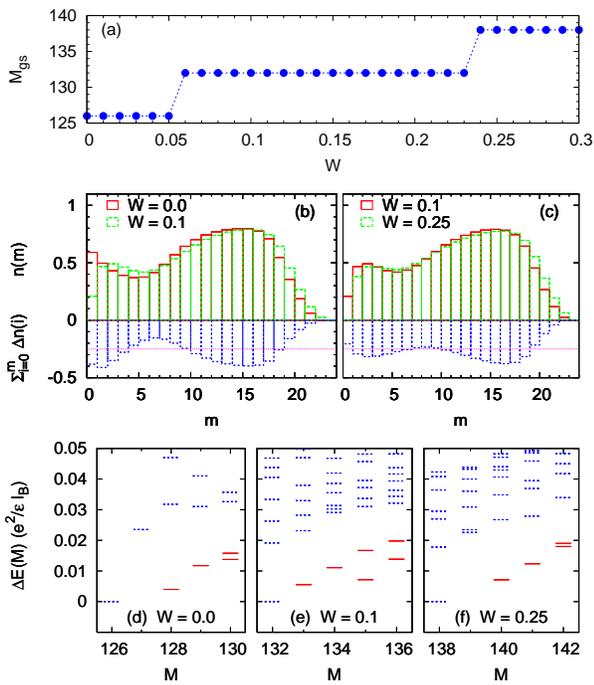}
\caption{ \label{fig:quasihole} (Color online) Generation of
quasiholes using a short-range repulsion $W c_0^{\dagger} c_0$
in a system of 12 electrons in 24 orbitals, for
the mixed Hamiltonian Eq. (\ref{eqn:mixedhamiltonian}) with
$\lambda=0.5$ and $d = 0.5 l_{\rm B}$. (a) Ground state angular
momentum $M_{\rm gs}$ as a function of $W$. (b) Electron occupation
number $n(m)$ of the ground state for $W = 0.0$ ($M_{\rm gs} = 126$)
and $W = 0.1$ ($M_{\rm gs} = 132$), as well as the accumulated
difference in $n(m)$ between the two states, $\sum_{i=0}^m \Delta
n(i)$, which oscillates around -0.25 (dotted line) for $m$ up to about 19,
indicating the emergence of a charge $+e/4$ quasihole at $m = 0$.
(c) $n(m)$ for $W = 0.1$ and $W = 0.25$ ($M_{\rm gs} = 138$), and
their accumulated difference, indicating the appearance of another
$+e/4$ quasihole. Low-energy edge excitations of the systems are
plotted for (d) $W = 0.0$, (e) $W = 0.1$, and (f) $W = 0.25$. The
fermionic mode supports 0, 1, 1, 2 states (solid bars) for $\Delta M
= 1$-4 in (d) ($M_{\rm gs} = 126$), but 1, 1, 2, 2 for $\Delta M = 1$-4
in (e) ($M_{\rm gs} = 132$). The numbers change back to 0, 1, 1, 2
for $\Delta M = 1$-4 in (f) ($M_{\rm gs} = 138$). 
This suggests that a single +e/4 quasihole (or an odd number of quasiholes)
changes the fermionic mode spectrum while a +e/2 quasihole (or, in general, an
even number of +e/4 quasiholes) does not.}
\end{figure}

{\it Charge $+e/4$ and $+e/2$ quasiholes.} One of the most important
properties of the MR state is that it supports charge $\pm e/4$
quasihole/particle excitations. To demonstrate the unusual
fractional charge, we add, to the mixed Hamiltonian with $\lambda =
0.5$ and $d = 0.5 l_{\rm B}$, a short-range potential: $H_W = W
c_0^{\dagger} c_0$, which tends to create quasiparticles or
quasiholes at the origin. For small enough repulsive $W$,
the ground state of the system should remain MR-like. As $W$ is
increased, a quasihole of charge $+e/4$ can appear at the origin,
reflected by a change of ground state angular momentum from
$M_{\rm gs} = N(2N-3)/2$ to $N(2N-3)/2 + N/2$, and depletion of $1/4$ in the
total occupation number of electrons at orbitals with small angular
momenta. If $W$ is increased further, a $+e/2$ quasihole, much like
a quasihole for the Laughlin state, appears near the origin in the
global ground state, whose total angular momentum further increases
to $N(2N-3)/2+N$. This is observed for a system of 12 electrons in
24 orbitals (as well as a smaller system of 10 electrons in 20 orbitals).
Figure~\ref{fig:quasihole}(a) shows the increase of
$M_{\rm gs}$ from 126 to 132 and then to 138 with increasing $W$.
Fig.~\ref{fig:quasihole}(b) compares the
electron occupation number $n(m)$ in each orbital for $W = 0.0$
($M_{\rm gs} = 126$, MR-like) and $W = 0.1$ ($M_{\rm gs} = 132$). The
accumulated difference in the occupation numbers of the two states,
$\sum_{i=0}^m \Delta n(i)$, oscillates around $-0.25$ for $m$ up to
about 19, indicating the existence of a $+e/4$ quasihole at the
origin. The same comparison for $W = 0.1$ ($M_{\rm gs} = 132$) and $W
= 0.25$ ($M_{\rm gs} = 138$) is plotted in
Fig.~\ref{fig:quasihole}(c). Their difference ($\sim -0.25$)
indicates the emergence of another $+e/4$ quasihole at the origin,
or a $+e/2$ quasihole compared to the MR-like state for $W = 0.0$.

The $+e/4$ quasihole supports a zero energy Majorana fermion mode
which is responsible for its non-Abelian nature. This zero mode
pairs with the edge excitations and changes their spectra. With
quasiholes in the bulk, the fermionic edge excitations are Majorana
fermions that obey either periodic (integer $l_{\rm f}$, twisted
sector) or antiperiodic (half integer $l_{\rm f}$, untwisted sector)
boundary conditions~\cite{milovanovic96}. In the presence of an odd
number of $+e/4$ quasiholes (twisted sector), two-fermion edge
excitations are shifted by $\delta(\Delta M) = -1$, relative to
those in the presence of an even number of such quasiholes
(untwisted sector). This is demonstrated in
Fig.~\ref{fig:quasihole}(d)-(f) for 12 electrons in 24 orbitals.
From $M_{\rm gs} = 126$ in Fig.~\ref{fig:quasihole}(d) ($W = 0.0$,
no quasihole present), we count the numbers of fermionic edge states
as 0, 1, 1, 2 for $\Delta M = 1$-4. From $M_{\rm gs} = 132$ in
Fig.~\ref{fig:quasihole}(e) ($W = 0.1$, one $+e/4$ quasihole
present), the numbers change to 1, 1, 2, 2 for $\Delta M = 1$-4. In
particular, the existence of a fermionic state at $\Delta M = 1$
clearly indicates the change. From $M_{\rm gs} = 138$ in
Fig.~\ref{fig:quasihole}(f) ($W = 0.25$, two $+e/4$ quasiholes
present), the numbers change back to 0, 1, 1, 2 for $\Delta M =
1$-4.

{\it Summary.} Our results suggest that the Moore-Read (MR) state
properly describes a half-filled first Landau level, for properly
chosen confinement potential. In this case the system supports
chiral edge excitations as well as fractionally charged quaisholes,
and their properties agree with theory predictions. We also find
that the window of stability of the MR state is rather narrow, and
the edge modes may suffer from reconstruction or other instabilities
as the confinement potential varies. The nature and consequences of
such instabilities are currently under investigation, which will be
presented elsewhere along with further details of the present work.

We thank Bert Halperin, Chetan Nayak, Nick Read, Zhenghan Wang and
Xiao-Gang Wen for very helpful discussions. This work is supported
by NSFC Project 10504028 (X.W.), and NSF grants No. DMR-0225698
(K.Y.) and No. DMR-0606566 (E.H.R.).


\begin{thebibliography}{99}
\bibitem{note} This is a simplification from the real system.
In reality the background charge equals the {\em total} electron
charge that includes the filled 0LL electrons; the latter are also
separated from the background charge, and the location of their edge
is different from that of the 1LL electrons. The consequences of
these will be discussed elsewhere.
\end{thebibliography}
\end{document}